\documentclass[a4paper]{article}

\usepackage[margin=1in]{geometry}

\usepackage[T1]{fontenc}
\newcommand{\changefont}[3]{
\fontfamily{#1} \fontseries{#2} \fontshape{#3} \selectfont}

\changefont{ptm}{m}{n}

\usepackage{setspace} 
 \usepackage{graphicx}    % needed for including graphics e.g. EPS, PS

\newcommand \be{\begin{equation}}
\newcommand \ee{\end{equation}}
\newcommand \ba{\begin{eqnarray}}
\newcommand \ea{\end{eqnarray}}

\def\bit{\begin{itemize}}
\def\eit{\end{itemize}}

\usepackage{amsfonts}
\usepackage{amssymb}
\usepackage{graphics}
\usepackage{mathrsfs}
\usepackage{color}

\long\def\symbolfootnote[#1]#2{\begingroup%
\def\thefootnote{\fnsymbol{footnote}}\footnote[#1]{#2}\endgroup} 

\begin{document}

%\begin{frontmatter}
%

\begin{center}
\Large \textbf{Entrainment by spatiotemporal chaos in glow discharge-semiconductor systems}
\end{center}

\begin{center}
\normalsize \textbf{Marat Akhmet$^{1,}\symbolfootnote[1]{Corresponding Author Tel.: +90 312 210 5355,  Fax: +90 312 210 2972, E-mail: marat@metu.edu.tr}$, Ismail  Rafatov$^2$, Mehmet Onur Fen$^1$} \\
\vspace{0.2cm}
\textit{\textbf{\footnotesize$^1$Department of Mathematics, Middle East Technical University, 06800, Ankara, Turkey,}}\\
\textit{\textbf{\footnotesize$^2$Department of Physics, Middle East Technical University, 06800, Ankara, Turkey}}
\vspace{0.1cm}
\vspace{0.1cm}
\end{center}

\vspace{0.3cm}

\begin{center}
\textbf{Abstract}
\end{center}

\noindent\ignorespaces

Entrainment of limit cycles by chaos \cite{Akh15} is discovered numerically through specially designed unidirectional coupling of two glow discharge-semiconductor systems. By utilizing the auxiliary system approach \cite{Abarbanel96}, it is verified that the phenomenon is not a chaos synchronization. Simulations demonstrate various aspects of the chaos appearance in both drive and response systems. Chaotic control is through the external circuit equation and governs the electrical potential on the boundary. The expandability of the theory to collectives of glow discharge systems is discussed, and this increases the potential of applications of the results. Moreover, the research completes the previous discussion of the chaos appearance in a glow discharge-semiconductor system \cite{Sijacic04}.

\vspace{0.2cm}
 
\noindent\ignorespaces \textbf{Keywords:} Glow discharge; Plasma; Spatiotemporal chaos; Entrainment by chaos; Period-doubling cascade; Cyclic chaos; Unidirectional coupling; Generalized synchronization.

\vspace{0.6cm}

\textbf{Spatiotemporal chaos is one of the complicated structures observed in spatially extended dynamical systems and it is characterized by chaotic properties both in time and space coordinates. The existence of a positive Lyapunov exponent can be used to detect spatiotemporal complexity, which can be observed, for example, in liquid crystal light valves, electroconvection, cardiac fibrillation, chemical reaction-diffusion systems and fluidized granular matter. Spatially extended dynamical systems often serve as standard models for the investigation of complex phenomena in electronics. A special interest is directed towards pattern-formation phenomena in electronic media, mainly the nonlinear gas discharge systems. It is clear that chaos can appear as an intrinsic property of systems as well as through couplings. The interaction of spatially extended systems is important for neural networks, reentry initiation in coupled parallel fibers, thermal convection in multilayered media and for systems consisting of several weakly coupled spatially extended systems such as the electrohydrodynamical convection in liquid crystals. In the present study, we numerically verify the appearance of cyclic irregular behavior (entrainment by chaos) in unidirectionally coupled glow discharge-semiconductor systems. The chaos in the response system is obtained through period-doubling cascade of the drive system such that it admits infinitely many unstable periodic solutions and sensitivity is present. Previously, the extension of chaos through couplings has been considered by synchronization \cite{Abarbanel96}, \cite{Rulkov95}-\cite{Pecora90}. The task is difficult for partial differential equations because of the choice of connecting parameters \cite{Kocarev97a}-\cite{Sushchik96}. Kocarev et al. \cite{Kocarev97a} suggested a useful time-discontinuous monitoring for synchronization, but our choice is based on a finite dimensional connection. It is demonstrated that the present results cannot be reduced to any one in the theory of synchronization of chaos. The technique of chaos extension suggested in the present research can be related to technical problems \cite{Kogelschatz02,Kogelschatz03}, where collectives of microdischarge systems  are considered and in models which appear in neural networks, hydrodynamics, optics, chemical reactions and electrical oscillators. Stabilization of multidimensional periodical regimes can be useful in applications of the glow discharge systems in conventional and energy saving lamps, beamers, flat TV screens, etc.}

%The technique used in this paper is as follows. We need a source of chaotic inputs, and for that reason we use a chaotic driving system. It is worth noting that the drive system is not necessarily needed in the chaotification procedure since it can be replaced by another source of a chaotic input, and in applications present result may be considered with, for example, chaotic inputs obtained from  experimental activity. So, initially, we take into account a chaotic glow discharge-semiconductor system (the drive system) and  use this system to influence in a unidirectional way another glow discharge-semiconductor system (the response) in such a manner that the response system mimics the same type of chaos of the drive system. In the present study, the chaos obtained through period-doubling cascade is used.   

%%%%%%%%%%%%%%%%%%%%%%%%%%%%%%%%%%%%%%%%%%%%%%%%%%%%%%%%%%%%%%%%%%%%%%%%%%%%%%%%%%%%%%%%%%%%%%%%%%%%%%%%%%%%%%%%%%%%%%%%%%%%%%%%%%%%%%%%%
%%%                                                       Introduction
%%%%%%%%%%%%%%%%%%%%%%%%%%%%%%%%%%%%%%%%%%%%%%%%%%%%%%%%%%%%%%%%%%%%%%%%%%%%%%%%%%%%%%%%%%%%%%%%%%%%%%%%%%%%%%%%%%%%%%%%%%%%%%%%%%%%%%%%%

\section{\label{sec:level0}Introduction}

The investigations of chaos theory for continuous-time dynamics started due to the needs of real world applications, especially with the studies of Poincar\'{e} \cite{Andersson94}, Cartwright and Littlewood \cite{Cartwright1}, Levinson \cite{Levinson}, Lorenz \cite{Lorenz63} and Ueda \cite{Ueda78}. Chaotic dynamics has high effectiveness in the analysis of electrical processes of neural networks \cite{Skarda87,Watanabe97} and can be used for optimization and self-organization problems in robotics \cite{Steingrube10}. The reason for that is the opportunities provided by the dynamical structure of chaos. 

Starting from the primary investigations \cite{Cartwright1}-\cite{Ueda78}, chaos has been found as an internal property of systems, and studies in this sense have prolonged until today, for example, by the construction of discrete maps \cite{Li75}-\cite{Akin03}. At the very beginning of the chaos analysis one has to mention the Smale Horseshoes technique \cite{Smale67} and symbolic dynamics \cite{Wiggins}. Another opportunity to reveal chaotic dynamics is the usage of bifurcation diagrams \cite{Coillet14,Cavalcante08}. 

% Our suggestions are based on the input/output mechanism which is a common technique in differential equations and previously used for obtaining bounded, periodic and almost periodic solutions. This time we realize the idea of mechanism for chaos generation. 

If one considers a mechanical or electrical system and perturb it by an external force which is bounded, periodic or almost periodic, then the forced system can produce a behavior with a similar property, boundedness/periodicity/almost periodicity \cite{Steinberg00}-\cite{Hagedorn}. A reasonable question appears whether it is possible to use a chaotic force to obtain the same type of irregularity in physical systems. 

To meet the challenge, we introduced rigorous description of chaotic force as a function or \textit{a set of functions} and described \textit{the input-output mechanism} for ordinary differential equations in the studies \cite{Akh15}, \cite{Akh2}-\cite{Akh18}. It was rigorously proved that an irregular behavior can follow the chaotic force very  likely as regular motions do. We have applied the machinery to mechanical and electrical systems with a finite number of freedom \cite{Akh15}, \cite{Akh7}-\cite{Akh16} as well as to neural networks \cite{Akh17,Akh18}. In the present study, we apply the theory to unidirectionally coupled glow discharge-semiconductor (GDS) systems.

\subsection{Preliminaries of the chaos extension}\label{intro_subsec1}

Chaotic dynamics can appear in systems as an intrinsic property and it can be extended through interactions. In the literature, an effective and unique way of the chaos extension from one system to another has been suggested within the scope of generalized synchronization \cite{Abarbanel96}, \cite{Rulkov95}-\cite{Gon04}, which characterizes the dynamics of a response system that is driven by the output of a chaotic driving system. Suppose that the dynamics of the drive and response systems are governed by the following systems with a skew product structure
\begin{eqnarray} \label{syncdrive}
x'=F(x)
\end{eqnarray}
and
\begin{eqnarray} \label{syncresponse}
y'=G(y,H(x)),
\end{eqnarray}
respectively, where $x\in\mathbb R^m, y\in\mathbb R^n.$ Generalized synchronization is said to occur if there exist sets $I_x,I_y$ of initial conditions and a transformation $\phi,$ defined on the chaotic attractor of (\ref{syncdrive}), such that for all $x(0)\in I_x$ and $y(0)\in I_y$ the relation $\displaystyle \lim_{t\to\infty} \left\|y(t)-\phi(x(t))\right\|=0$ holds. In that case, a motion which starts on $I_x\times I_y$ collapses onto a manifold $M\subset I_x\times I_y$ of synchronized motions. The transformation $\phi$ is not required to exist for the transient trajectories. When $\phi$ is the identity, the identical synchronization takes place \cite{Gon04,Pecora90}.

The synchronization of a large class of unidirectionally coupled chaotic partial differential equations was deeply investigated in \cite{Kocarev97a,Kocarev97b}, where the synchronization was achieved by applying the driving signals only at a finite number of space points. The synchronization of spatiotemporal chaos in a pair of complex Ginzburg-Landau equations was performed in \cite{Sushchik96} for the case when all space points are continuously driven. In the present study, we use perturbations to a single coordinate of an infinite dimensional response system, which is non-chaotic in the absence of driving, to obtain chaotic motions in the system.

It has not been investigated whether the response system admits the same type of chaos with the drive system in the theory of chaos synchronization yet. The replication of chaos with specific types such as Devaney \cite{Dev90}, Li-Yorke \cite{Li75} and period-doubling cascade \cite{Feigenbaum80}-\cite{Sander13} was investigated for drive-response couples for the first time in our papers \cite{Akh15}, \cite{Akh2}-\cite{Akh18}.

In the study \cite{Akh15}, we considered a system of the form 
\begin{eqnarray} \label{systemwithcycle}
u'=K(u),
\end{eqnarray}
where $K:\mathbb R^n \to \mathbb R^n$ is a continuously differentiable function. We supposed that system (\ref{systemwithcycle}) possesses an orbitally stable limit cycle and perturbed it with solutions of a chaos generating system, in the form of (\ref{syncdrive}), and set up the system
\begin{eqnarray}\label{responsesystem2}
y'=K(y)+\mu M(x),
\end{eqnarray}
where $\mu$ is a nonzero number and $M:\mathbb R^m \to \mathbb R^n$ is a continuous function. The extension of sensitivity and chaos through period-doubling cascade for the coupled system (\ref{syncdrive})-(\ref{responsesystem2}) were rigorously proved in the paper \cite{Akh15}. As a result, we achieved \textit{chaotic cycles}, that is, motions which behave cyclically and chaotically, simultaneously.

%The results obtained in the paper \cite{Akh15} cannot be reduced to generalized synchronization. This chaotification method was also used in the paper \cite{Akh18} to obtain cyclic irregular behavior in neural networks, which is important in neuroscience \cite{Skarda87}. In this paper, we demonstrate that our results cannot be reduced to generalized synchronization at all. In the case that chaos is attracted by equilibria \cite{Akh8,Akh16,Akh17}, one can show that the response system admits the same ingredients of chaos of the drive system as well as the asymptotic closeness of the solutions of the drive and response systems is true for our approach. However, in the theory of synchronization \cite{Rulkov95}-\cite{Pecora90}, the ingredients of chaos, such as sensitivity and existence of infinitely many unstable periodic solutions, are not considered. In \cite{Akh15,Akh16}, the attraction of chaos in the response system is verified, and  by the auxiliary system approach \cite{Abarbanel96} and evaluation of conditional Lyapunov exponents \cite{Kocarev96,Pecora90} it is shown that the results cannot be explained by generalized synchronization.  

The rich experience of chaos expansion in finite dimensional spaces provides a confidence that our approach mentioned in \cite{Akh15} has to work also in infinite dimensional spaces. In this paper, we numerically observe the presence of orbitally stable limit cycles in the $2-$dimensional projections of the infinite dimensional space as well as their deformation to chaotic cycles under chaotic perturbations. By using the technique presented in \cite{Akh15}, one can elaborate the results of the present study from the theoretical point of view. Although couplings of GDS systems have not been performed in the literature yet, our results reveal the opportunity of chaos extension in such systems.
 
Summarizing, electronic systems are important tools for synchronization and chaos extension. In this paper, we make use of our previous approach \cite{Akh15} to extend chaos in unidirectionally coupled GDS systems. 

% We show the deformation of limit cycles to chaotic cycles in response GDS systems numerically.

\subsection{Description of the GDS system model}\label{intro_subsec2} 
\begin{figure}[ht]
\begin{center}
\includegraphics[width=7.0cm]{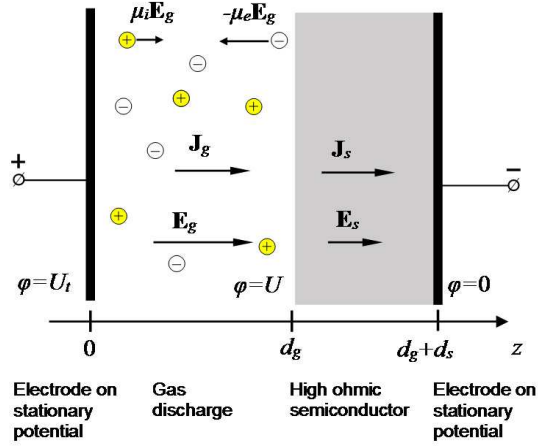} \\
\caption{A cross section of a planar discharge cell: it consist of a metal anode, a gas layer, a high-ohmic cathode, and another metal contact. The subscripts $g$ and $s$ refer to the gas and semiconductor regions.}
\label{gas_discharge}
\end{center}
\end{figure}

Our GDS was previously studied both theoretically and experimentally in \cite{Sijacic04}, \cite{Sijacic05}-\cite{Gurevich03}. It represents a planar plasma layer coupled to a planar semiconductor layer, which are sandwiched between two planar electrodes to which a DC voltage is applied (see Fig. \ref{gas_discharge}). We used one-dimensional fluid model for this system, where any pattern formation in the transversal direction is excluded and only the single dimension normal to the layers is resolved. For the gas discharge, the model takes into account electron and ion drift in the electric field, bulk impact ionization and secondary emission from the cathode as well as space charge effects. The semiconductor is approximated with a constant conductivity. 

The gas-discharge part of the model consists of continuity equations for two charged species, namely,
electrons and positive ions with particle densities $n_{\rm e}$ and $n_{i}$:
\begin{eqnarray} \label{pde1}
\partial{_t}n_{\rm e}+\nabla\cdot{\bf \Gamma}_{\rm e} &=& {\it S_e},
\label{eq:c1}\\
\partial{_t}n_{i}+\nabla\cdot{\bf \Gamma}_{i} &=& {\it S_i},
\label{eq:c2}
\end{eqnarray}
which are coupled to Poisson's equation for the electric field in electrostatic approximation:
\begin{equation}
\nabla\cdot{\bf E}=\frac{e}{\varepsilon_{0}}\left(n_{i}-n_{\rm
e}\right), ~~ {\bf E}=-\nabla{\Phi}.
\end{equation}
Here, $\Phi$ is the electric potential, ${\bf E}$ is the electric field in the gas discharge, $e$ is the
elementary charge, and $\varepsilon_{0}$ is the dielectric constant. The vector fields ${\bf \Gamma}_{\rm e}$
and ${\bf \Gamma}_{i}$ are the particle flux densities, that in simplest approximation are described by
drift only. (In general, particle diffusion $D_{e,i}\nabla n_{e,i}$ could be included.) The drift velocities are assumed to
depend linearly on the local electric field with mobilities $\mu_{e}\gg \mu_{i}$:
\begin{eqnarray}
{\bf \Gamma}_{\rm e} = -\mu_{\rm e}n_{\rm e}\,{\bf E},~~~ {\bf \Gamma}_{i} = \mu_{i}n_{i}\,{\bf E},
\end{eqnarray}
hence the total electric current in the discharge is
\begin{equation} \label{pde5}
\label{eq:Je} {\bf J}= \epsilon_0\partial_t{\bf E}+ e\,\big({\bf \Gamma}_i-{\bf \Gamma}_{\rm e}\big)=\epsilon_0\partial_t{\bf E}+e\,\Big(\mu_in_i+\mu_{\rm e}n_{\rm
e}\Big)\,{\bf E}.
\end{equation}
Two types of ionization processes are taken into account: the $\alpha$ process of electron impact ionization
in the bulk of the gas, and the $\gamma$ process of electron emission by ion impact onto the cathode. In a
local field approximation, the $\alpha$ process determines the source terms in the continuity equations
(\ref{eq:c1}) and (\ref{eq:c2}):
\begin{equation}
{\it S_{e}}={\it S_{i}}=|{\bf \Gamma}_{\rm e}|\,\alpha_{0}\,\alpha\left({|{\bf E}|}/{E_0}\right),
\end{equation}
where we use the classical Townsend approximation
\begin{equation}
\alpha\left(|{\bf E}|/E_{0}\right)=\exp\left(-E{_0}/|{\bf
E}|\right).
\end{equation}
The effect of the semiconductor layer with thickness $d_s$, conductivity $\sigma_s$, dielectric constant $\varepsilon_s$   is described by the external circuit equation
\begin{eqnarray}\label{ece} 
\partial_t U=\frac{U_t-U-R_sJ}{T_s},
\end{eqnarray}
where $U_t$ is the applied voltage, $U=\int_0 ^{d_g}E\,dZ$ is the voltage over the gas discharge which is the electric field $E$ integrated over the height $d_g$ of the discharge, $R_s=d_s/\sigma_s$ is the resistance of the semiconductor layer, where $\sigma_s$ is its conductivity, and $T_s=\epsilon_s\epsilon_0/\sigma_s$ is the Maxwell relaxation time of the semiconductor with dielectric constant $\varepsilon_s$.

%\textbf{In the present study, we investigate the dynamics of coupled GDS systems. For this reason, we apply the parameters  $n_{\rm e},$  $n_{i},$ $\Phi,$  ${\bf E},$ subject to the system of partial differential equations (\ref{pde1})-(\ref{pde5}) and $U,$ which satisfies the ordinary differential equation (\ref{ece}), with the applied voltage $U_t$  \cite{Sijacic04,Sijacic05}. }

Following the traditions of the synchronization of chaotic systems, we will call the coupled GDS systems as the \textit{drive} and \textit{response} systems.
 
The goal of our investigation is to extend the spatiotemporal chaos of a drive GDS system to a response GDS system by means of a special connection mechanism between the systems. In order to make our present study self-sufficient, we complete the chaos analysis of the GDS system, which was initiated in the papers \cite{Sijacic04,Sijacic05}. The method of the analysis, as well as the connection mechanism are our theoretical suggestions \cite{Akh15}, \cite{Akh2}-\cite{Akh18} and in the present circumstances one can say that we consider entrainment by chaos \cite{Akh15} for GDS systems. Entrainment by chaos, which  is the deformation of limit cycles to chaotic cycles, is verified by simulations. 
 
The chaos obtained through period-doubling cascade \cite{Feigenbaum80}-\cite{Sander13} is under investigation in the present study. In other words, the existence of infinitely many unstable periodic solutions and the presence of sensitivity \cite{Dev90} are considered. One of the advantages of our approach is the controllability of the extended chaos \cite{Akh15,Gon04,Akh7,Akh8,Pyragas92}. It is possible to stabilize an unstable periodic solution of the response GDS system by controlling the chaos of the drive system. The presented technique is applicable to large number interconnected GDS systems and the control of the global chaos can also be achieved. This approach can be useful for applications of the gas discharge systems in conventional and energy saving lamps, beamers, flat TV screens, etc. \cite{Kogelschatz02,Kogelschatz03}.

%%%%%%%%%%%%%%%%%%%%%

\subsection{Formulation of the model in dimensionless form}

The dimensional analysis is performed essentially as in \cite{Sijacic04,Sijacic05}. In dimensional units, $Z$ parametrizes the direction normal to the layers. The anode of the gas discharge is at $Z=0$, the cathode end of the discharge is at $Z=d_g$, and the semiconductor extends up to $Z=d_g+d_s$.

When diffusion is neglected, the ion current and the ion density at the anode vanish. This is described by
the boundary condition on the anode $Z=0$:
\begin{eqnarray}
\label{BC0} {\bf \Gamma}_{i}\left(0,t\right)=0 ~~~\Rightarrow~~~ n_{i}\left(0,t\right)=0.
\end{eqnarray}
The boundary condition at the cathode, $Z=d_g$, describes the $\gamma$-process of secondary electron
emission:
\begin{eqnarray}
\label{BCdg}
\left|{\bf \Gamma}_{\rm e}\left(d_g,t\right)\right|
=\gamma\left|{\bf \Gamma}_{i}\left(d_g,t\right)\right|     
~~~ \Rightarrow~~~ \mu_{\rm e} n_{\rm e}\left(d_g,t\right)= \gamma
\mu_{i}n_{i}\left(d_g,t\right).
\end{eqnarray}
Finally, a DC voltage $U_{t}$ is applied to the system determining the electric potential on the boundaries
\begin{eqnarray}
\Phi\left(0,t\right)=0,~~~
\Phi\left(d_{g},t\right)=-U.
\end{eqnarray}
Here, the first potential vanishes due to gauge freedom.

Let us introduce the intrinsic parameters of the system as $\displaystyle t_{0}=\frac{1}{\alpha_{0}\mu_{e}E_{0}},$ $\displaystyle Z_{0}=\frac{1}{\alpha_{0}},$ $\displaystyle n_{0}=\frac{\varepsilon_{0}\alpha_{0}E_{0}}{e}.$  In the studies \cite{Sijacic04,Sijacic05}, the problem was reduced to one spatial dimension $z$ such that the GDS system takes the following dimensionless form,
\begin{eqnarray}\label{pde_eqn1}
\begin{array}{l}
\partial_\tau\sigma-\partial_z\left(\cal{E}\sigma\right)=\sigma\cal{E}\alpha\left(\cal{E}\right),\\  
\partial_\tau\rho+\mu\partial_z\left(\cal{E}\rho\right)=\sigma\cal{E}\alpha\left(\cal{E}\right),\\  
\partial_z{\cal{E}}= \rho-\sigma, ~~ {\cal{E}}=-\partial_z\phi, 
\end{array}
\end{eqnarray}
where the dimensionless time, coordinates and fields are  $\displaystyle {z}=\frac{{Z}}{Z_0},$  $\displaystyle \tau=\frac{t}{t_0},$  $\displaystyle \sigma({z},\tau)=\frac{n_{\rm e}\left({Z},t\right)}{n_{0}},$  $\displaystyle \rho({z},\tau)=\frac{n_{i}\left({Z},t\right)}{n_{0}},$  $\displaystyle {\cal{E}}({z},\tau)=\frac{{E}\left({Z},t\right)}{E_{0}},$  $\displaystyle \phi({z},\tau)=\frac{\Phi\left({Z},t\right)}{E_0 Z_0}$ and $\displaystyle \alpha\left(\cal{E}\right)=e^{-1/|{\cal{E}}|}.$

The intrinsic dimensionless parameters of the gas discharge are the mobility ratio $\mu$ of electrons and ions and the length ratio $L$ of discharge gap width and impact ionization length. That is, $\displaystyle \mu=\frac{\mu_{i}}{\mu_{e}}$ and $L=\displaystyle \frac{d_{g}}{Z_{0}}.$ The boundary conditions become
\begin{eqnarray}\label{pde_eqn1_bc}
\begin{array}{l}
\rho(0,\tau)=0,\\
\sigma(L,\tau)=\gamma\mu\rho(L,\tau),\\
\phi(0,\tau)=0, ~~ \phi(L,\tau)=-{\cal{U}},
\end{array}
\end{eqnarray}
and the external circuit is described by
\begin{eqnarray} \label{pde_main_system}
\partial_\tau {\cal{U}}=\frac{{\cal{U}}_t-{\cal{U}}-{\cal{R}}_sj}{\tau_s},
\end{eqnarray}
where the total applied voltage is rescaled as ${\cal U}_{t}={U_{t}}/({E_{0}Z_{0}})$, dimensionless voltage ${\cal{U}}(\tau)=\int_0 ^{L}{\cal{E}}\,dz$, time scale $\tau_s=T_s/t_0$, resistance ${\cal{R}}_s=R_s\,e\mu_en_0/Z_0$, and spatially conserved total current $j(\tau)=\partial_{\tau}{\cal{E}}+\mu\rho{\cal{E}}+\sigma{\cal{E}}$.

We consider a regime corresponding to a transition between Townsend and glow discharge. The parameters are taken as in the experiments \cite{Str2} and in our previous work \cite{Sijacic04}.
The discharge is in nitrogen at 40 mbar, in a gap of 1.4 mm. We used the ion mobility $\mu_{i}=23.33 ~{\rm cm^{2}/(V \, s)}$ and electron mobility $\mu_{e}=6666.6 ~{\rm cm^{2}/(V \, s)}$, therefore the mobility ratio is $\mu=\mu_{i}/\mu_{e}=0.0035$. The secondary emission coefficient
was taken as $\gamma=0.08$. The applied voltages $U_t$ are in the range of 513--570 V. For $\alpha_0=Ap=[27.8\;\mu{\rm m}]^{-1}$ and for $E_0=Bp=10.3\;{\rm kV/cm}$, we used values from \cite{Raizer}. The semiconductor layer consists of 1.5 mm
of GaAs with dielectric constant $\varepsilon_s=13.1$ and conductivity $\sigma_s=(2.6\times 10^5 ~\Omega \,{\rm cm})^{-1}$. Corresponding dimensionless parameters are $L=50$, ${\cal{R}}_s=30597$, $\tau_s=7435$, and a total voltage range ${\cal{U}}_t$  between $17.67$ and $20.03.$

\section{Chaotically coupled GDS systems}
 
In the present section, we will extend the spatiotemporal chaos of a drive GDS system through utilizing its voltage over the gas discharge as a chaotic control applied to the electric circuit of a response GDS system. In the coupling, the voltage over the discharge of the drive system is applied as a perturbation to the  circuit equation of the response system. The presence of entrainment by chaos in the response system will be shown numerically. Moreover, we will compare our results with generalized synchronization.

The full analysis of the spatiotemporal chaos in the GDS system (\ref{pde_eqn1}-\ref{pde_main_system}) is provided in the Appendix, where the bifurcation diagram  as well as the chaotic behaviors in the voltage, electric field, electron density and ion density of the system are represented. According to these results,
the GDS system
\begin{eqnarray} \label{generator_pde}
\begin{array}{l}
\partial_\tau\sigma-\partial_z\left(\cal{E}\sigma\right)=\sigma\cal{E}\alpha\left(\cal{E}\right),\\
\partial_\tau\rho+\mu\partial_z\left(\cal{E}\rho\right)=\sigma\cal{E}\alpha\left(\cal{E}\right),\\
\partial_z{\cal{E}}= \rho-\sigma, ~~ {\cal{E}}=-\partial_z\phi,\\
\displaystyle \partial_\tau {\cal{U}}=\frac{20-{\cal{U}}-{\cal{R}}_sj}{\tau_s},
\end{array}
\end{eqnarray}
is chaotic, and it will be accompanied by the boundary conditions
\begin{eqnarray*} 
&& \rho(0,\tau)=0,\\
&& \sigma(L,\tau)=\gamma\mu\rho(L,\tau),\\
&& \phi(0,\tau)=0, ~~ \phi(L,\tau)=-{\cal{U}}.
\end{eqnarray*}
We will take into account (\ref{generator_pde}) as the drive system.

The solutions of (\ref{generator_pde}) will be used as a perturbation for the response GDS system in the form,
\begin{eqnarray} \label{replicator_pde}
\begin{array}{l}
\partial_\tau\widetilde{\sigma}-\partial_z\left(\widetilde{\cal{E}}\widetilde{\sigma}\right)=\widetilde{\sigma}\widetilde{\cal{E}}\alpha\left(\widetilde{\cal{E}}\right),\\
 \partial_{\tau}\widetilde{\rho}+\mu\partial_z\left(\widetilde{\cal{E}}\widetilde{\rho}\right)=\widetilde{\sigma}\widetilde{\cal{E}}\alpha\left(\widetilde{\cal{E}}\right),\\
 \partial_z{\widetilde{\cal{E}}}= \widetilde{\rho}-\widetilde{\sigma}, ~~ {\widetilde{\cal{E}}}=-\partial_z\widetilde{\phi},\\
\displaystyle \partial_\tau {\cal{V}}=\frac{{\cal V}_t- {\cal V} -{\cal{R}}_s\widetilde{j}+\delta{\cal{U}}(\tau)}{\tau_s},
\end{array}
\end{eqnarray}
with the boundary conditions
\begin{eqnarray*} 
&& \widetilde{\rho}(0,\tau)=0,\\
&& \widetilde{\sigma}(L,\tau)=\gamma\mu\widetilde{\rho}(L,\tau),\\
&& \widetilde{\phi}(0,\tau)=0, ~~ \widetilde{\phi}(L,\tau)=-{\cal{V}}. 
\end{eqnarray*}
In system (\ref{replicator_pde}), $\delta$ is a nonzero number and the term $\delta{\cal U}(\tau)/ \tau_{s}$ is the perturbation from the drive system (\ref{generator_pde}).

It is shown in the Appendix for the parameter value ${\cal U}_t=17.7$ that the projection of the attractor of  system (\ref{pde_eqn1})-(\ref{pde_main_system}) on the domain of equation (\ref{pde_main_system}) is a stable limit cycle (see Fig. \ref{stable_attractor}). That is, in the absence of driving, the response system (\ref{replicator_pde}) with ${\cal V}_t=17.7$ does not possess chaos. We will numerically show that the response GDS system possesses chaotic motions near the limit cycle, provided that the driving effect is included. Our results are theoretically based on the study \cite{Akh15}, where we have proved that if the drive system admits infinitely many unstable periodic solutions and sensitivity, then the response system does the same. Since the attractor exists in system (\ref{pde_eqn1})-(\ref{pde_main_system}) with ${\cal U}_t=17.7$, one can conclude by the extension of our results presented in \cite{Akh15} that if the number $|\delta|$ in equation (\ref{replicator_pde}) is sufficiently small, then system (\ref{replicator_pde}) possesses cyclic chaos. That is, entrainment by chaos takes place in the system.

\begin{figure}[ht]
\begin{center}
\includegraphics[width=6.5cm]{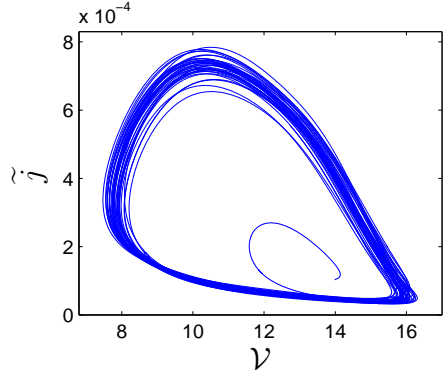} \\
\caption{The trajectory of the response system (\ref{replicator_pde}) in the ${\cal V}-\widetilde j$ plane manifests the chaotic cycle.}
\label{response_attractor}
\end{center}
\end{figure}

\begin{figure}[ht]
\begin{center}
\includegraphics[width=6.5cm]{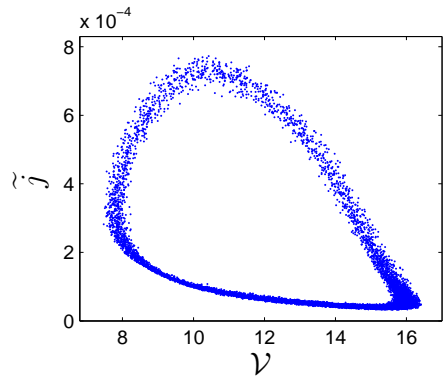} \\
\caption{The projection of the stroboscopic plot of system (\ref{generator_pde})-(\ref{replicator_pde})  on the ${\cal V}-\widetilde j$ plane reveals the presence of chaotic behavior in the response system.}
\label{response_stroboscopic}
\end{center}
\end{figure}

\begin{figure}[ht]
\begin{center}
\includegraphics[width=11.0cm]{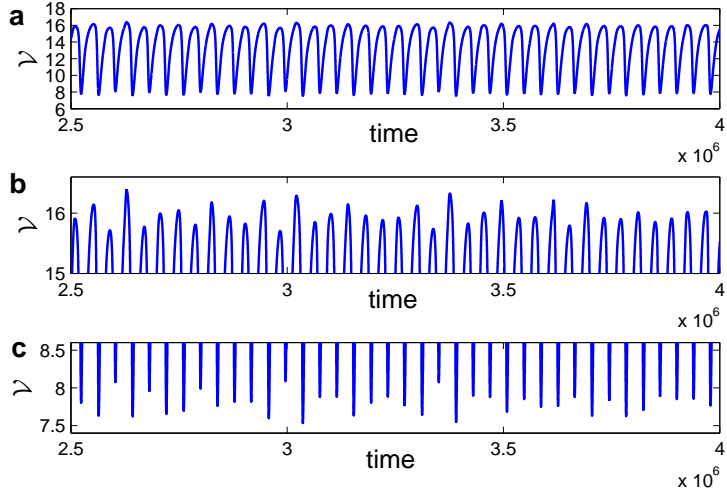} \\
\caption{The behavior of the $\cal V$ coordinate of system (\ref{replicator_pde}) is shown in (a). In (b) and (c), where the chaotic behavior is observable, the amplitudes are restricted to the ranges $15-16.6$ and $7.4-8.6,$ respectively.}
\label{response_timeseries}
\end{center}
\end{figure}

Let us take ${\cal V}_t=17.7$ and  $\delta=0.047$ in the response GDS system (\ref{replicator_pde}). Using the solution of the drive system shown in Figures \ref{drive_attractor}, \ref{drive_timeseries} and \ref{drive3D}, we depict in Figure \ref{response_attractor}  the projection of a chaotic solution of (\ref{replicator_pde}) on the ${\cal V}-\widetilde{j}$ plane. On the other hand, the projection of the stroboscopic plot of the response system  on the same plane is shown in Figure \ref{response_stroboscopic}. Both of the figures reveal that the response GDS system possesses motions that behave chaotically around the limit cycle of system (\ref{pde_eqn1})-(\ref{pde_main_system}) with ${\cal U}_t=17.7.$ Moreover, to support the presence of chaos in the response system, we depict in Figure \ref{response_timeseries} the time series of the $\cal V$ coordinate. The amplitude ranges $15-16.6$ and $7.4-8.6$ are used in Figure \ref{response_timeseries}, (b) and (c), respectively, to increase the visibility of chaotic behavior.
 
\begin{figure}[ht!]
\begin{center}
\includegraphics[width=8.45cm]{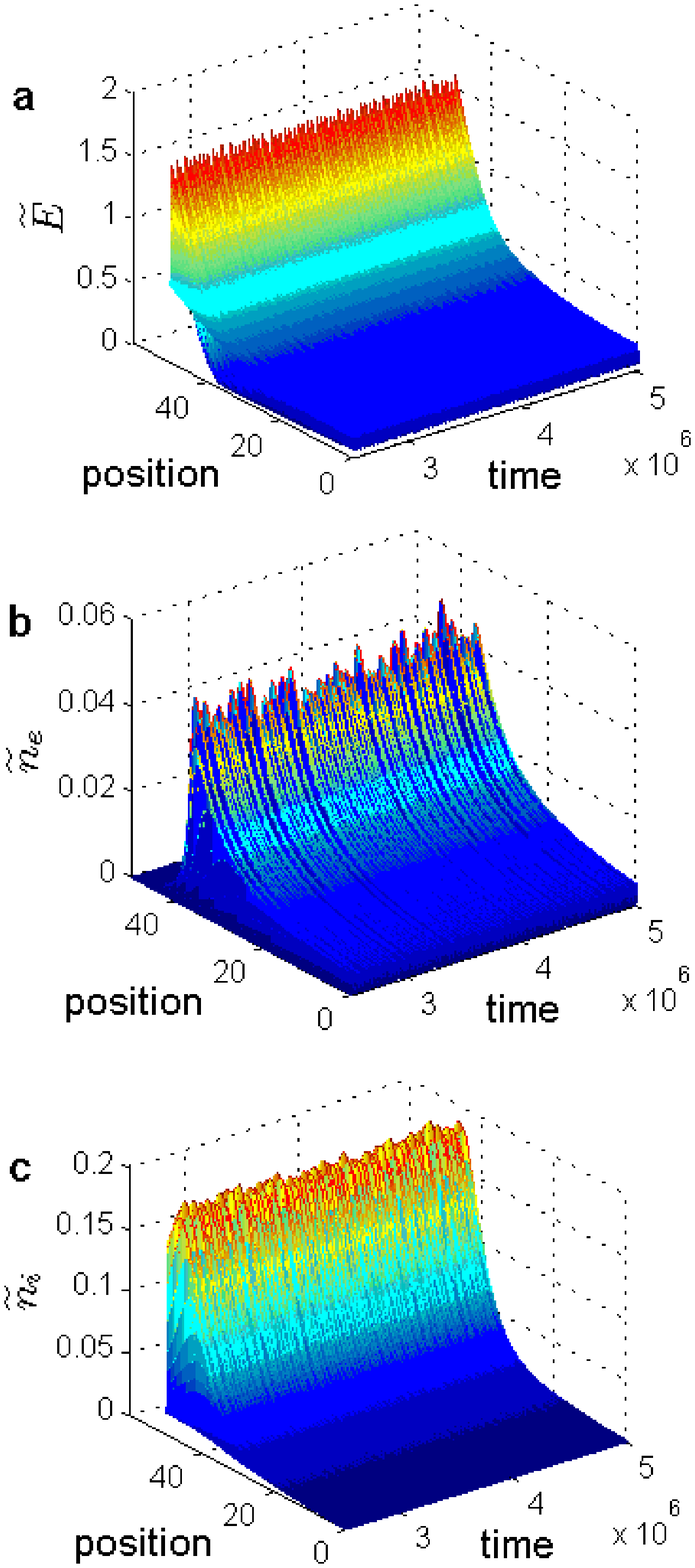} \\
\caption{Time evolution of profiles of the (a) electric field $\widetilde{E},$ (b) electron
density $\widetilde{n}_e,$ and (c) ion density $\widetilde{n}_i$ support the existence of chaotic
motions around the periodic solution.}
\label{response3D}
\end{center}
\end{figure}

Figure \ref{response3D}, (a), (b) and (c) depict, respectively, the chaotic behavior in the electric field, electron density and ion density of system  (\ref{replicator_pde}). These figures also support the presence of motions that behave chaotically around the limit cycle.

Now, let us compare our results with generalized synchronization (GS) \cite{Rulkov95}-\cite{Gon04}. According to Kocarev and Parlitz (1996), GS occurs for the coupled systems (\ref{syncdrive}) and (\ref{syncresponse}) if and only if for all $x_0\in I_x,$ $y_{10},$ $y_{20}\in I_y,$ the  asymptotic stability criterion  
$\displaystyle \lim_{t\to\infty} \left\| y(t,x_0,y_{10}) - y(t,x_0,y_{20})  \right\|=0$ holds, where $y(t,x_0,y_{10})$ and $y(t,x_0,y_{20})$ denote the solutions of (\ref{syncresponse}) with the initial data $y(0,x_0,y_{10})=y_{10},$ $y(0,x_0,y_{20})=y_{20}$ and the same $x(t),$ $x(0)=x_0.$ This criterion is a mathematical formulation of the auxiliary system approach \cite{Abarbanel96,Gon04}.  We shall make use of the auxiliary system approach  to demonstrate the absence of generalized synchronization in the coupled system (\ref{generator_pde})-(\ref{replicator_pde}). 
 
We introduce the auxiliary system  
\begin{eqnarray} \label{auxiliary_pde}
\begin{array}{l}
\partial_\tau\overline{\sigma}-\partial_z\left(\overline{\cal{E}}\overline{\sigma}\right)=\overline{\sigma}\overline{\cal{E}}\alpha\left(\overline{\cal{E}}\right),\\
\partial_{\tau}\overline{\rho}+\mu\partial_z\left(\overline{\cal{E}}\overline{\rho}\right)=\overline{\sigma}\overline{\cal{E}}\alpha\left(\overline{\cal{E}}\right),\\
 \partial_z{\overline{\cal{E}}}= \widetilde{\rho}-\overline{\sigma}, ~~ {\overline{\cal{E}}}=-\partial_z\overline{\phi},\\
\displaystyle \partial_\tau {\cal{W}}=\frac{17.7-{\cal{W}}-{\cal{R}}_s\overline{j}+0.047\cal{U}(\tau)}{\tau_s}
\end{array}
\end{eqnarray}
with the boundary conditions
\begin{eqnarray*}
&& \overline{\rho}(0,\tau)=0,\\
&& \overline{\sigma}(L,\tau)=\gamma\mu\overline{\rho}(L,\tau),\\
&& \overline{\phi}(0,\tau)=0, ~~ \overline{\phi}(L,\tau)=-{\cal{W}}. 
\end{eqnarray*} 
 
\begin{figure}[ht]
\begin{center}
\includegraphics[width=6.5cm]{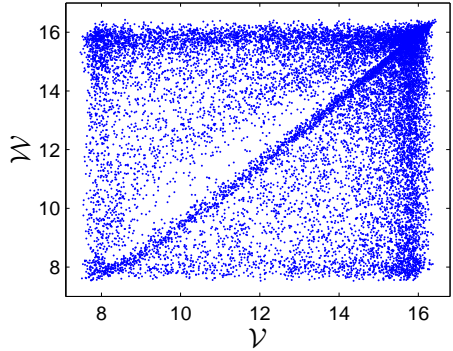} \\
\caption{Application of the auxiliary system approach reveals that the coupled systems (\ref{generator_pde}) and (\ref{replicator_pde}) are not synchronized.}
\label{aux_system}
\end{center}
\end{figure}

Making use of the solution ${\cal U(\tau)}$ whose graph is represented in Figure \ref{drive_timeseries} in both of the systems (\ref{replicator_pde}) and (\ref{auxiliary_pde}), we depict in Figure \ref{aux_system} the projection of the stroboscopic plot of system (\ref{replicator_pde})-(\ref{auxiliary_pde}) on the ${\cal V}-{\cal W}$ plane. The first $500$ iterations are omitted in the simulation. The time interval $[0,80\times 10^6]$ is used and the time step is taken as $\Delta \tau=5000.$ Since the plot does not take place on the line ${\cal W}={\cal V},$ we conclude that \textit{generalized synchronization is not achieved} in the dynamics of the coupled system (\ref{generator_pde})-(\ref{replicator_pde}).

\section{Conclusions}

In the studies \cite{Akh15}, \cite{Akh2}-\cite{Akh18}, we applied the input-output mechanism to systems that admit stable equilibrium points as well as limit cycles. It is theoretically proved in \cite{Akh15} that weak forcing of systems with stable limit cycles leads to the deformation of limit cycles to chaotic cycles, that is motions that behave chaotically around the limit cycle. This phenomenon, which is called \textit{entrainment by chaos}, cannot be explained by the theory of generalized synchronization \cite{Rulkov95}-\cite{Gon04}, and it is also used in the present study. In the electrical sense, the chaotification of limit cycles is much more preferable than that procedure for asymptotic equilibria, because of the role of oscillations for electronics. Accordingly, in this paper, we demonstrate the entrainment by chaos in coupled glow discharge  systems.

%The analysis of chaos for GDS systems that was started in the papers \cite{Sijacic04,Sijacic05} is completed in the present study, and the extension of chaotic behavior is demonstrated by establishing unidirectional coupling between such systems. 

In this paper, we utilize GDS systems as drive and response electrical models. GDS systems were analyzed for a chaos presence in \cite{Sijacic04}. We complete the analysis by constructing the full period-doubling bifurcation diagram to demonstrate that the drive system admits infinitely many unstable periodic solutions as well as sensitivity. However, this is only an auxiliary result. The main novelty of the present article with respect to the previous studies \cite{Sijacic04,Sijacic05,Rafatov07} is that we consider these systems which are coupled in a unidirectional way and prove that the chaos can be extended through  couplings of GDS systems as well as in their arbitrary large collectives. This type of chaos extension may give benefits in further applications, for example, in economic lamps and flat TV screens \cite{Kogelschatz02,Kogelschatz03}. We suggest that our way of numerical analysis and special design of complexity can be further verified experimentally. It is worth noting that our approach is not generalized synchronization of chaos at all. This is demonstrated through the special method of auxiliary system approach \cite{Abarbanel96,Gon04}.

%\textbf{The dynamics of a response system that is driven by the output of a chaotic driving system can be characterized by the concept of generalized synchronization. This type of synchronization occurs if there exists a functional relation between the states of the drive and response systems. The presence of generalized synchronization allows us to determine completely the dynamics of the response by the dynamics of the drive \cite{Rulkov95}-\cite{Gon04}. One can take into account generalized synchronization as a tool for chaos generation in the response system by applying chaos of the drive system as a perturbation. In the present study, we introduce another way for the indication of chaos in the response system not through the asymptotic properties of solutions, but directly by using the chaotic ingredients or the process of chaos appearance with respect to the parameter changes. To emphasize the difference, we call such type of chaos generation as \textit{replication of chaos}. The important point in the theoretical discussion is the existence of an attractor for the response system in the absence of driving. We proved that these attractors can be stable equilibria, limit cycles or tori \cite{Akh8}-\cite{Akh18}. The theoretical results combined with simulations help us to obtain chaos in unidirectionally coupled glow discharge-semiconductor systems. This is the main result of our study, and its mathematical background depends on the paper \cite{Akh15}. }

\section*{Acknowledgments}
I. Rafatov acknowledges support by the Scientific and Technical Research Council of Turkey (TUBITAK), research grant 212T164.

\section*{Appendix: The chaos in the drive GDS system}

In this part, we will extend the results of \cite{Sijacic04} about the presence of chaos in GDS systems. In the article \cite{Sijacic04}, only a finite number of period-doubling bifurcations were indicated. However, in the present study, we represent the occurrence of infinitely many period-doubling bifurcations by means of a bifurcation diagram and we definitely reveal the regions of regularity and chaoticity. 

\begin{figure}[ht]
\begin{center}
\includegraphics[width=0.75\textwidth]{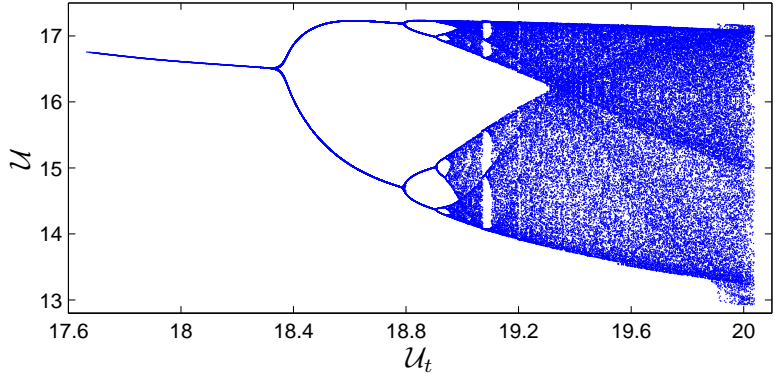} \\
\caption{The bifurcation diagram of system (\ref{pde_eqn1})-(\ref{pde_main_system}) for the values of the parameter ${\cal U}_t$ between $17.67$ and $20.03$.}
\label{pde_bif_diagram}
\end{center}
\end{figure}

The bifurcation diagram corresponding to the ${\cal U}$ coordinate of system (\ref{pde_eqn1})-(\ref{pde_main_system}) with the boundary conditions (\ref{pde_eqn1_bc}) is pictured in Figure \ref{pde_bif_diagram}. Here, ${\cal U}_t$ is the bifurcation parameter. Supporting the results of \cite{Sijacic04}, it is observable in the figure that the system displays period-doubling bifurcations and leads to chaos. The period-doubling bifurcations occur approximately at the ${\cal U}_t$ values $18.315,$ $18.782,$ $18.902,$ $18.939,$ etc., and a period-six window appears near ${\cal U}_t=19.073$ in the bifurcation diagram.   

\begin{figure}[ht]
\begin{center}
\includegraphics[width=6.5cm]{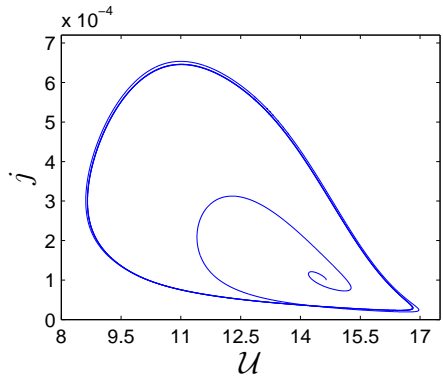} \\
\caption{The figure reveals   a limit cycle, the projection of the attractor of the global system  on the domain of equation (\ref{pde_main_system}) with ${\cal U}_t=17.7.$ }
\label{stable_attractor}
\end{center}
\end{figure}

One can conclude from the bifurcation diagram that the system (\ref{pde_eqn1})-(\ref{pde_main_system}) possesses a stable periodic solution for ${\cal U}_t=17.7.$ 
The projection of a solution that approaches to the stable limit cycle, which is the projection of the attractor of the global system (\ref{pde_eqn1})-(\ref{pde_main_system}) on the domain of (\ref{pde_main_system}) with ${\cal U}_t=17.7,$ is depicted in Figure \ref{stable_attractor}. This result confirms the existence of an attractor as a periodic solution in the spatiotemporal equation.

\begin{figure}[ht]
\begin{center}
\includegraphics[width=6.5cm]{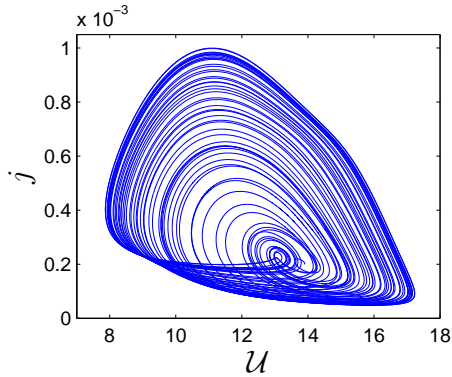} \\
\caption{The projection of the chaotic solution of the drive GDS system (\ref{generator_pde}) on the ${\cal U}-j$ plane.}
\label{drive_attractor}
\end{center}
\end{figure}

\begin{figure}[ht]
\begin{center}
\includegraphics[width=12.5cm]{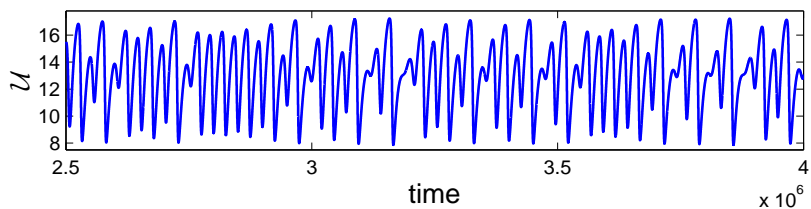} \\
\caption{The chaotic behavior of the $\cal U$ coordinate of system (\ref{generator_pde}).}
\label{drive_timeseries}
\end{center}
\end{figure}

The bifurcation diagram shown in Figure \ref{pde_bif_diagram} confirms that the drive GDS system (\ref{generator_pde}) is chaotic. The projection of a chaotic solution of (\ref{generator_pde}) on the ${\cal U}-j$ plane is represented in Figure \ref{drive_attractor}. Moreover, the time series of the ${\cal U}$ coordinate of the same solution is shown Figure \ref{drive_timeseries}, where one can see the chaotic behavior.

\begin{figure}[ht!]
\begin{center}
\includegraphics[width=8.45cm]{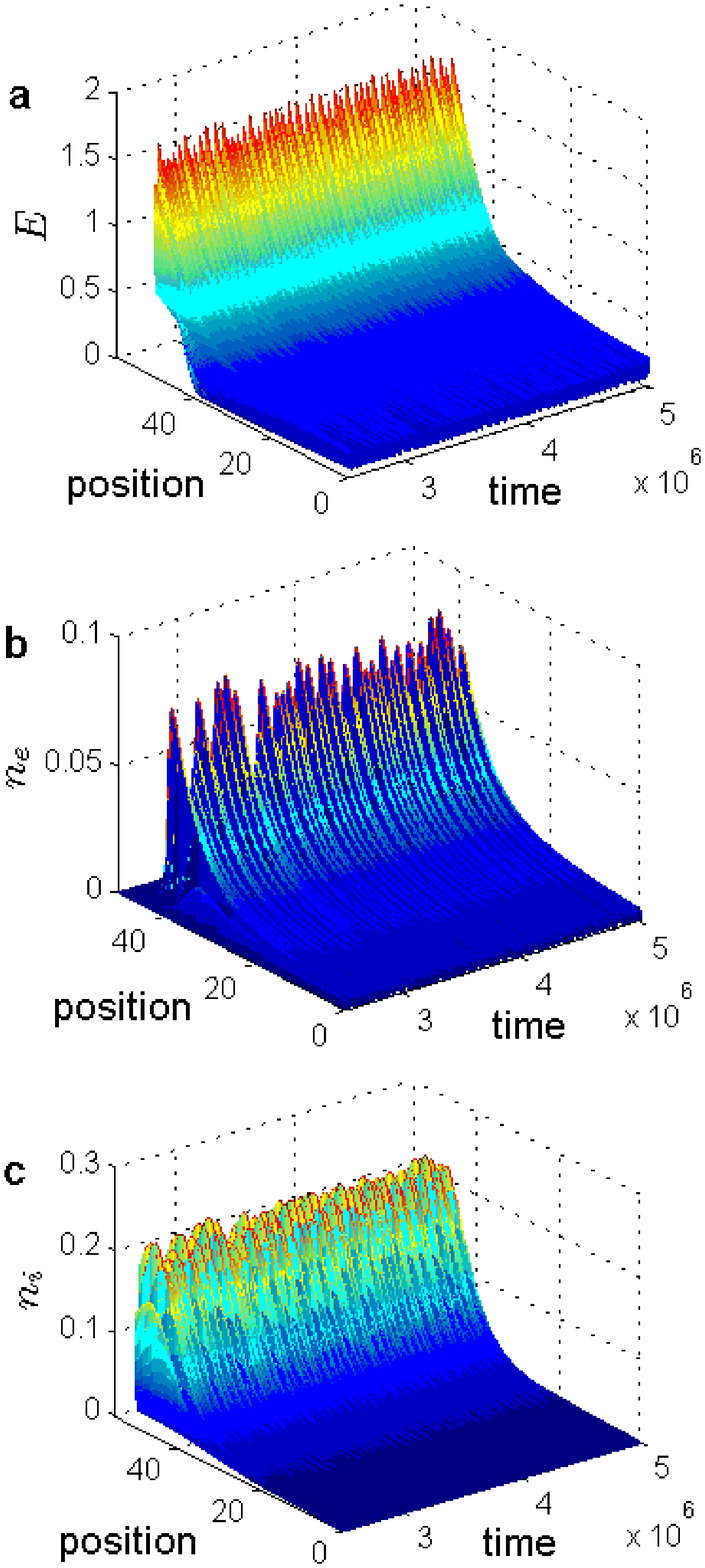} \\ 
\caption{Profiles of the (a) electric field $E,$ (b) electron density $n_e,$ and (c) ion density $n_i$ as functions of time support the existence of spatiotemporal chaos.}
\label{drive3D}
\end{center}
\end{figure}

The profiles of the electric field $E$, electron density $n_{\rm e}$ and ion density $n_i$ of (\ref{generator_pde}) are pictured in Figure \ref{drive3D}, (a), (b) and (c), respectively. Figure \ref{drive3D} also confirms the presence of chaos in the drive system.

%\newpage
%
%\begin{center}  \large\textbf{Figure Captions} \end{center}
%\vspace{.4cm}
%
%
%\textbf{Figure 1.} The chaotic behavior of the $SICNNs$ (\ref{ex1}).
%
%\vspace{.4cm}
%
%\textbf{Figure 2.} The chaotic behavior of the $SICNNs$ (\ref{ex2}). 
%
%\vspace{.4cm}
%
%\textbf{Figure 3.} The projection of the chaotic attractor of the network (\ref{ex2}) on the $x_{22}-x_{31}-x_{33}$ space.

\end{document}